\documentclass[aip,pof,preprint,numerical]{revtex4-2}

\usepackage{graphicx}
\usepackage{amssymb}
\usepackage{amsmath}
\usepackage{booktabs}
\usepackage{longtable}
\usepackage{bm} 
\usepackage[utf8]{inputenc}
\usepackage[T1]{fontenc}
\usepackage{mathptmx}
\usepackage{etoolbox}
\usepackage{todonotes}

\let\vec\mathbf

\def\ti{\mathrm{i}}
\def\td{\mathrm{d}}

\usepackage[colorlinks=true, allcolors=blue]{hyperref}

\draft 

\begin{document}


\title{Lees-Edwards boundary conditions for translation invariant shear flow:\\implementation and transport properties} 

\author{Sebastian Bindgen}
\thanks{ORCID:\protect\href{https://orcid.org/0000-0002-5861-0990}{0000-0002-5861-0990}}
\affiliation{KU Leuven, Soft Matter, Rheology and Technology, Celestijnenlaan 200f - box 2424, 3001 Leuven, Belgium}
\author{Florian Weik}
\author{Rudolf Weeber}
\affiliation{University of Stuttgart, Institute for Computational Physics,  Allmandring 3, 70569 Stuttgart, Germany}
\author{Erin Koos}
\thanks{ORCID:\protect\href{https://orcid.org/0000-0002-2468-2312}{0000-0002-2468-2312}}
\affiliation{KU Leuven, Soft Matter, Rheology and Technology, Celestijnenlaan 200f - box 2424, 3001 Leuven, Belgium}
\author{Pierre de Buyl}
\thanks{ORCID:\protect\href{http://orcid.org/0000-0002-6640-6463}{0000-0002-6640-6463}}
\email[corresponding author:]{pierre.debuyl@kuleuven.be}
\affiliation{Royal Meteorological Institute of Belgium, Avenue Circulaire 3, 1180 Brussels, Belgium}
\affiliation{KU Leuven, Institute for Theoretical Physics, Celestijnenlaan 200d - box 2415, 3001 Leuven, Belgium}

\date{\today}

\begin{abstract}
Molecular dynamics (MD) simulations represent a powerful investigation tool in the field of soft
matter.
By using shear flows, one can probe the bulk rheology of complex fluids, also beyond the
linear response regime, in a way that imitates laboratory experiments.
One solution to impose a shear flow in particle-based simulations is the Lees-Edwards
technique which ensures that particles experience shear by imposing rules for motion and
interactions across the boundary in the direction of the shear plane.
Despite their presentation in 1972, a readily available public implementation of
Lees-Edwards boundary conditions has been missing from MD simulation codes.
In this article, we present our implementation of the Lees-Edwards technique and discuss the
relevant technical choices. We used ESPResSo, the extensible simulation package for research
on soft matter, for molecular dynamics simulations which can be used as a reference for
other implementers.
We illustrate our implementation using bulk dissipative particle dynamics fluids, compare
different viscosity measurement techniques, and observe the anomalous diffusion in our
samples during continuous and oscillatory shear, in good comparison to theoretical
estimates.
\end{abstract}


\maketitle 

\section{Introduction}

The response of many soft matter materials to deformations includes a viscous and an elastic
component, thereby giving rise to the category of viscoelastic materials.
Many experimental techniques have been developed throughout the years to investigate these
unique properties, in both the linear and non-linear regime, including the combination of
confocal microscope imaging and other techniques to monitor microstructural changes \cite{Smith1724}.
Despite these new developments, many experimental challenges remain \cite{Wilson2011},
such as probing shorter time and length scales.
Optical \cite{Tassieri2019} and magnetic tweezers \cite{Rich2011} enable the monitoring of
minimal forces and displacements, but remain restricted to small observation windows.

Computer simulations represent an interesting alternative to experimental observations for
soft condensed matter.
In Molecular Dynamics (MD) simulations, the movement of atoms is governed by Newton's laws
of motion. One can thus access the coordinates of the atoms and compute all relevant
observables~\cite{Frenkel2002,Allen2017}.
The size and duration of MD simulations is mainly restricted by the available computational
power. Dedicated computational methods have been developed to address larger system sizes
and durations with respect to atomistic methods, such as the coarse-graining of atoms into
``effective atoms'' that represent atomic ensembles (e.g. water molecules or polymeric units), colloidal particles, or fluid elements. For the latter, methods include techniques such as dissipative particle dynamics (DPD)\cite{Hoogerbrugge1992} and multi-particle collision dynamics MPCD.\cite{malevanets1999, kapral2008} Another important feature that needs to be considered is the interaction of e.g. colloidal particles with their surroundings.\cite{Lee2008}

Thermostats can also be used to achieve a set temperature, as opposed to constant-energy
systems, using for instance the Langevin thermostat whose properties are well
known~\cite{kubo_fluctuation-dissipation_1966}.
In practice, the thermostatting is achieved by adding a random force, the noise, and a
dissipative force, the friction, whose magnitudes are related by the fluctuation dissipation
theorem.
Such simulations, using a Langevin thermostat, do not represent the fluid flows and can only
mimic liquid-like behavior.
More specifically, they do not take collective effects into account and do not conserve
momentum.  Recent developments related to the fluctuation dissipation theorem could solve
some of these restrictions. This allows the linear modulus of more complex systems,
including yield stress systems, to be accessible.\cite{Wittmer2015}
Despite these improvements, the fact that the Langevin thermostat breaks momentum
conservation makes it a poor choice for pseudorheological measurements.

The DPD method was introduced for the simulation of
thermostatted particle-based soft-matter systems~\cite{Hoogerbrugge1992,Espanol1995,Soddemann2003}.
DPD simulations use only pairwise forces, including for noise and friction, are not
restricted to the linear regime, and provide direct access to nonequilibrium situations for
which Green-Kubo methods would not apply.
Due to momentum conservation, DPD can be used to study hydrodynamic phenomena.
Applications of DPD include polymer solutions \cite{Symeonidis2005}, colloidal
suspensions \cite{Pan2009}, multiphase flow phenomena \cite{Pan2014} and biological systems
\cite{Li2013}. Recent investigations \cite{Leimkuhler2016} have shown further ways to
improve the accuracy of the DPD method and pointed out its possible use to investigate
non-linear material behaviour.

To simulate the non-linear behavior using pseudorheological measurements, several solutions are available. The simplest way to introduce a deformation of the simulated volume element is to abandon the periodicity in the shear plane. Replacing these boundaries with moving walls leads to a shear flow \cite{Khare1996}. Applications of such simulations can be found in, e.g., the migration of polymer in small gaps as present in bearings \cite{Kreer2001}.
The drawback of this method is the loss in periodicity and the appearance of boundary
effects. Large systems can only be simulated by increasing the size of the primary
simulations box, which entails a corresponding increase of computational work load.
Two interesting techniques have been developed to avoid using walls for driving the shear
flow.
The first is the SLLOD technique, which consists of modified equations of motion in which
the flow velocity is added to the particles' motion~\cite{Edberg1987, Travis1995,
  Evans2008}.
When using the SLLOD equations of motion, thermostatting is done on the peculiar velocity of
the particles (the laboratory reference frame minus the assigned flow velocity). In
practice, this means that the flow profile is imposed via a bias in the
thermostat~\cite{Shang2017, Imperio2008, Hess2002}. As the flow velocity is different at the
boundaries of the shear plane, the simulation boundaries must be adjusted by deforming the
box according to the shear velocity.
Furthermore, there are several possibilities to implement the SLLOD method with differing
theoretical backgrounds. There is no consensus on which implementation should be preferred
and on the limitations that each one entails. These discussions include the limitations of
each single technique in term of applicable general flow patterns and the need for
artificial external forces.\cite{edwards2005, daivis2006,edwards2006}
The SLLOD technique is available in the LAMMPS package (Large-scale Atomic/Molecular
Massively Parallel Simulator~\cite{plimpton1995}), for instance.
The second method consists in establishing, via an applied periodic external force, a
periodic flow profile~\cite{Hess2002}. The periodic flow method is convenient to implement
as it does not require modification to the box geometry or to the boundary conditions.
Whereas the results of SLLOD simulation can be used to simulate materials in a simple shear
flow, they do not provide any feedback between the material structure and the flow profile,
which means they fail with regard to more complex systems such as yield stress
fluids~\cite{Hess2002}. Shear banding is one of the many effects that cannot be observed
with such a technique \cite{Cao2012}. Furthermore, it is not possible to achieve correct
hydrodynamics, since the bulk fluid is only modelled implicitly. Viscous losses cannot be
measured and hence the loss modulus of any material is inaccessible.
A limitation of periodic flow simulations is that one cannot use them for linear shear
profiles.

The most promising technique to combine the advantages of the previous mentioned approaches
are Lees-Edwards boundary conditions (LEbc). Introduced in 1972 \cite{Lees1972}, they are a
technique to address non-linear material behaviour during flow, and are distinguished from
other non-linear simulation methods as they do not require a biased thermostat
\cite{Evans1986, Evans2008} or non-periodic walls to initiate a shear flow, but rely on the
specific rules at the boundary that lead to a translationally invariant system. LEbc are
sometimes referred to as sliding brick boundary conditions \cite{Leimkuhler2016}. Despite
being developed more than four decades ago, there is presently no open-source implementation
of the Lees-Edwards boundary condition.
There is a clear need for such flow phenomena simulations and simulations using the
Lees-Edwards boundary conditions are of broad interest in academic as well as industrial
research.

In this paper, we present the principle of the Lees-Edwards method and its implementation in
the ESPResSo molecular simulation package in section~\ref{sec:implementation}.
We provide the corresponding code under the same open-source license as ESPResSo. The source
code availability, as well as the parameter and analysis files, are discussed in appendix
\ref{sec:reproc}.
We describe the simulation methods, including the details of the dissipative particle
dynamics method, in section~\ref{sec:methods}.
We present our results on the self-diffusion of DPD particles and on the viscosity of the
DPD fluid in section~\ref{sec:results} and conclude in section~\ref{sec:conclusions}.

\section{Lees-Edwards boundary conditions}
\label{sec:implementation}

\subsection{Principle of Lees Edwards boundary conditions} \label{subsec:LE}

Lees-Edwards boundary conditions (LEbc), are a generalisation of the periodic boundary conditions for systems undergoing
shear \cite{Lees1972}.
With periodic boundary conditions, a particle exiting the simulation cell is replaced at its
periodic location inside the cell and the computation of distances across the boundaries
uses the minimal image convention~\cite{Allen2017}.
When using LEbc, a particle crossing the shear plane is also replaced in the simulation
box. The position and velocity of the particle, however, are shifted so that the trajectory
of the particle is compatible with its image in the adjacent moving cell.
The LEbc thus allows the simulation of infinitely extended systems, as with periodic
boundary conditions with a prescribed shear, using a finite simulation cell. 

In the stationary regime, a constant shear flow in the shear direction, here the $x$-direction, is obtained and the simulation has translational invariance in the direction normal to the shear plane, i.e., the gradient ($y$) and vorticity ($z$) direction. The change in position $x'$ as a function of time $t$ for a particle that leaves the computational domain in the velocity gradient direction normal to the shear plane is
\begin{equation}
x'(t) = x(t) + x_{\text{LE}}
\end{equation}
where the Lees-Edwards offset, $x_{\text{LE}}$ the displacement of the adjacent simulation
cell with respect to the primary cell, is
\begin{equation}
x_{\text{LE}} = v_{\text{LE}} \cdot t
\end{equation}
for steady shear, with $v_{\text{LE}}$ as the Lees-Edwards velocity. The change in velocity is
\begin{equation}
v_x'(t) = v_x(t) + v_{\text{LE}}
\end{equation}
based on the drift velocity $v_{\text{LE}}$ of the periodic images. This can be seen in Figure~\ref{fig:integrator}(a)
where a particle leaves the primary simulation box and is re-introduced at position $p''$ instead of $p'$.
The updated position is then wrapped into the primary simulation cell.
When the periodic boundary conditions in the other directions remain unaltered these modifications result in a shear flow of the magnitude $\dot\gamma = v_\text{LE} / h$, where $h$ is the height of the simulation box (in $y$).

\subsection{Application of Lees-Edwards boundary conditions to the velocity Verlet integrator}

To implement the principle of the Lees-Edwards boundary conditions in a molecular dynamics
(MD) program, it is necessary to specify practical details: the computation of distances, of
relative velocities, and the combination with the velocity Verlet algorithm
\cite{Verlet1967, Swope1982}.
We implemented the LEbc method in the ESPResSo package with the goal to provide a
reference implementation of LEbc and a user friendly interface for steady shear and for
sinusoidal shear, which is useful to determine the dynamic moduli $G'$ and $G''$.

The update of a particle's coordinates in the velocity Verlet integrator occurs in the
following order:
\begin{align}
  v^\ast &= v(t) + \frac{1}{2 m} f(t)   \tag{VV 1} \label{vv1}  \times \Delta t \\
  x(t+\Delta t) &= x(t) + \frac{1}{2} \left( v(t) + v^\ast \right) \times \Delta t \tag{VV 2} \label{vv2} \\
  &\textrm{update all forces at time } t+\Delta t \textrm{ , using } x(t+\Delta t)\cr
  v(t+\Delta t) &= v^\ast + \frac{1}{2 m} f(t+\Delta t) \tag{VV 3}\label{vv3} \times \Delta t
\end{align}
where $\Delta t$ is the time step, $m$ is the particle mass, and $f$ is the force on the particle.

\begin{figure*}[htbp]
\begin{center}
\includegraphics[width=\textwidth]{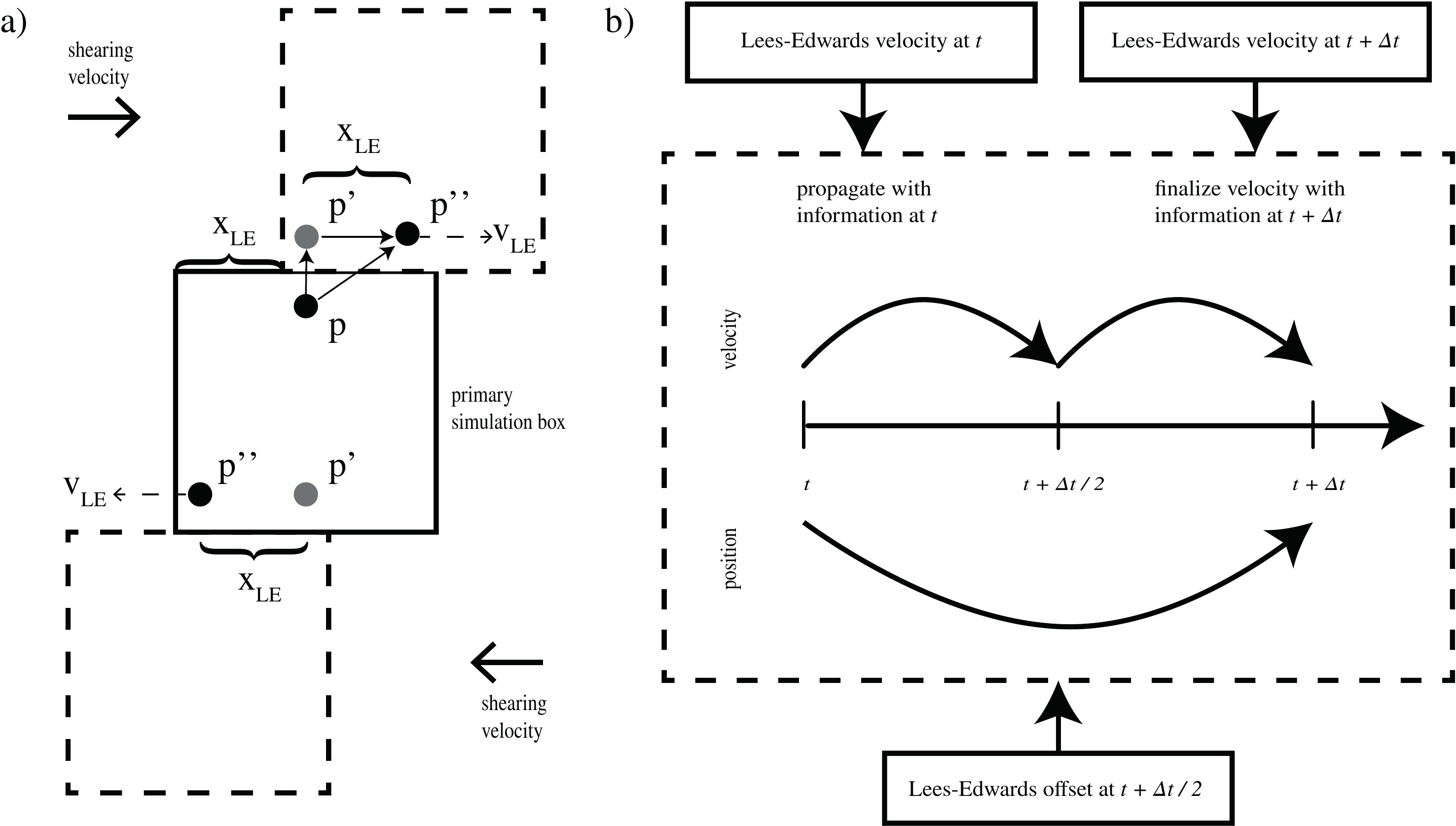}
\caption{{\bf
(a) A simplified representation of the particle movement including the Lees-Edwards boundary conditions. The particle is reintroduced at position $p''$ instead of $p'$ due to the shift in the image boxes. This has to be captured in the distance function as indicated by the arrows crossing the box boundary.
(b) Changes to the commonly used velocity Verlet integrator, which is displayed in the central box introduced with the inclusion of the Lees-Edwards boundary conditions. As indicated by the surrounding boxes, the Lees-Edwards velocity is updated at $t$ and $t+\Delta t$, and the offset is included at $t+\Delta t /2 $.
}}
\label{fig:integrator}
\end{center}
\end{figure*}

To translate the LEbc, we must apply the rules exposed above within this framework, which
leads to a two step approach for the application of the velocity and position jumps which we
illustrate in Figure~\ref{fig:integrator}(b). A one step approach for the integration is not suitable as it can lead
to numerical instability.
In the first step, we determine the Lees-Edwards velocity at the present simulation time $t$,
i.e. $v_{\text{LE}}(t)$ and the Lees-Edwards offset at one half time-step $\Delta t$ ahead
of the simulation time: $x_{\text{LE}}(t+ \frac{\Delta t}{2})$.
After the position update in step \eqref{vv2}, we check if the particle has left the
primary computational domain $0 \leq y(t+\Delta t) < h$. In that case, we apply the
position jump
\begin{equation}
x(t+\Delta t) \to x(t+\Delta t) - x_\text{LE}\left(t + \frac{\Delta t}{2}\right)
\end{equation}
and apply half the velocity jump
\begin{equation}
v^{\ast} \to v^\ast - v_\text{LE}(t) / 2 ~,
\end{equation}
following the structure of the velocity Verlet scheme.
The forces, including the DPD dissipative
and random forces, are updated at the middle of the time step of the integration. 
It is thus necessary to have current velocities to ensure the correct
thermalisation of the particles. Therefore, we tag the particle as undergoing the Lees-Edwards transformation, 
so that we can apply the second part of the jump after step
\eqref{vv3}, using $v_\text{LE}(t+\Delta t)$.

As simulation cells move with respect to each other due to the shear, the distance
and the relative velocity between particles across the shear plane must include the
Less-Edwards offset. This is shown in Figure~\ref{fig:integrator}(a) where the arrows 
crossing the boundary represent the applied distance function. While particles might
experience a positive offset while crossing the box, the distance function in fact must
include the negative offset to reflect these changes correctly. 
Accordingly, we modified the distance function in ESPResSo, so that the appropriate offset distance
is used for all actions, such as computing the force, as well as building the neighbour lists.
The modification of the relative velocities is necessary for the computation of the
velocity-dependent DPD forces. Thus the velocity difference function is also modified in ESPResSo.

The trajectories in LEbc simulations will display discontinuous jumps whenever a particle crosses the
shear plane. Since LEbc simulations include an infinitely extended system, it
is possible to reconstruct physically meaningful trajectories. For this purpose, we store the
accumulated offset in position during the numerical integration $x_{\text{LE}}$ of a
particle and its movement in a periodic image $i(t)$, along the shear direction, according to
\begin{equation}
\label{eq:accumulated-offset}
x_{\text{part, LE}} = \sum_j x_{\text{LE}}(t_{\text{jump}}) + \sum_t \Delta t \cdot v_{\text{LE}}(t) \cdot i(t).
\end{equation}
where $j$ stands for occurring jumps at the LE boundary.
$x_{\text{part, LE}}$ represents the displacement of the particle as it moves outside of the
primary simulation cell.
This data, which is necessary for the reconstruction of the full trajectory of the particle,
must be computed as the simulation proceeds and cannot be obtained later using only recorded
positions and velocities.

\subsection{Modification of the cell system}
\label{sec:cell-system}

In principle, the number of pairs in a system of $N$ particles is $\mathcal{O}(N^2)$. Such a
high computational cost of the force calculation is unpractical. It is, in principle,
possible to compute only $\mathcal{O}(N)$ pair forces as long as only short-range forces are
used, which can be cut off after a certain distance.
This can be accomplished via neighbor lists and it is often practical to sort the particles
into cells. This is realised in ESPResSo with the technique of domain decomposition, where
the system is partitioned into cubic cells for the purpose of storing the particles'
coordinates and for spatially sorting the particles\cite{Smith1991}.
The sliding nature of the boundary in shear flow simulations breaks the periodic assumption
on which the domain decomposition is based and requires an appropriate modification.

\begin{figure}[htbp]
\begin{center}
\includegraphics[width=0.5\textwidth]{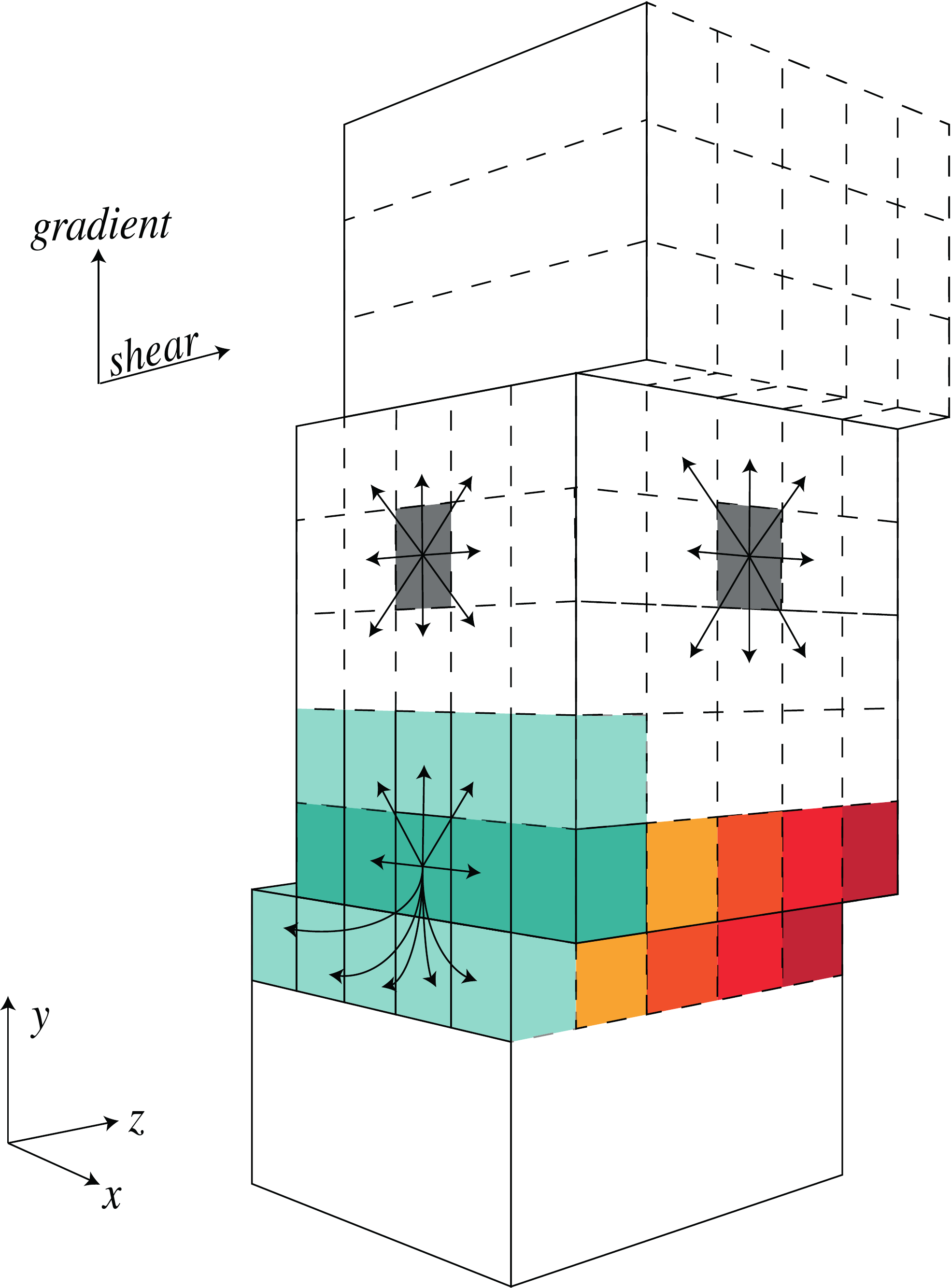}
\caption{\bf{Communication pattern for the columnar cell system. The arrows shown on the $x$-$y$ surface show how the local cells communicate with all other cells of the same $x$-column and $y \pm 1$. For local cells in the other directions the usual domain decomposition method is used which is shown here.}}
\label{fig:columnarcellsystem}
\end{center}
\end{figure}

To keep the computational advantage of domain decomposition, we introduce a
columnar domain decomposition: we treat all cells in the layer adjacent to the boundary of
the primary simulation box as neighbours, as shown using the orange-red colors in Figure 2.
It does not influence the domain decomposition in the gradient and vorticity directions.  A
special node grid that consist of [x, y, z] = [m, n, o], i.e.
$N_{\text{nodes}} = m \cdot n \cdot o$ nodes, has to be chosen. This grid must be chosen such that it has
exactly one node in the shear direction,
i.e. the x-direction as shear direction leads to a [1, n, o] node grid. This guarantees
that no possible particle interactions are lost or considered twice due to the Lees-Edwards
offset. In this way, a ``re-wiring'' of the cell-neighborship relations during a running simulation can be avoided.
Figure \ref{fig:columnarcellsystem} shows a representation of this system and illustrates
the column as well as the communication directions used.

The cells located near the Lees-Edwards boundary, oriented along a column in the shear direction, communicate with all other cells in the $x$-column as well as with the cells in the columns directly above and below ($y \pm 1$). Thus, as shown in the $x$-$y$ plane wit arrows, all teal cells are considered in the neighbor list. These interactions are superimposed onto the usual domain decomposition between local cells, as shown for the grey box on the $y$-$z$ plane with the limitation that all cells in a column communicate with each other. 
Communications in the shear directions are carried out via all cells in a column, where as for communications in the other two directions a regular domain decomposition is used. During all distance calculations the modified minimum image vector is used because it accounts for jumps across the box.
The strategy chosen here minimizes the changes to the source code of the existing simulation
package ESPResSo.

\subsection{Spurious discontinuity of the velocity profile}

Some articles about the Lees-Edwards method have reported discontinuous velocity profiles near
the shear plane boundary.
We consider here the work of Chatterjee \cite{Chatterjee2007}: in order to mitigate the
discontinuity in the velocity profile, Chatterjee proposed to disable the thermostatting
(pairwise friction and noise terms) in the vicinity of the simulation cell boundary that
corresponds to the shear plane.
Since the LEbc method is invariant under translation, Leimkuhler and Shang argued that the
strategy of Chatterjee was only necessary in order to counteract programming errors in the
simulation code~\cite{Leimkuhler2016}. In their article, Leimkuhler and Shang verify their
hypothesis by introducing voluntarily the suspected bug in their own simulation code. In the
presence of this bug, they are able to reproduce the suspicious profile observed by
Chatterjee.
In our implementation, we have not introduced any local change to the thermostat and
find a continuous and linear velocity profile for the DPD particles. We present those
results in Section~\ref{subsec:flowprofile}.

\section{Simulation methods}
\label{sec:methods}

We implemented the Lees-Edwards method, from scratch, within the software package ESPResSo
(Extensible Simulation Package for Research on Soft Matter) \cite{Arnold2013a, Arnold2013b,
  Limbach2006, Weik2019}.
We provide the details on the specific version in appendix~\ref{sec:reproc}.

\subsection{Dissipative particle dynamics}

We consider a bulk fluid consisting of point particles with a mass $m$, a position $\vec{r}_\ti$ and a
velocity $\vec{v}_\ti$. We use Molecular Dynamics (MD), which solves Newton's equation for
all particles, subject to interaction forces:
\begin{equation}\label{newton}
m \frac{\td^2 \vec{r}_\ti}{\td t^2} = \vec{f}_\ti
\end{equation}
where $\vec{f}_\text{i}$ is the total force on the i-th particle.
For dissipative particle dynamics (DPD) \cite{groot1997}, there are two modifications from this starting
point: the particles are coarse-grained to represent effective fluid elements instead of
atoms, and the relative velocity of particle pairs is thermostatted to introduce thermal motion and damping.
The pairwise thermostatting method in DPD implies that the method conserves linear momentum
and can be used for hydrodynamic simulations, which sets it apart from Langevin dynamics,
another common choice in coarse-grained MD simulations.

We follow the presentation of DPD by Groot and Warren\cite{groot1997}, to which we refer the
readers for more background about the method.
The total DPD force $\vec{f}^{\textrm{DPD}}_\text{i}$ on particle $i$ can be decomposed into:
\begin{equation}
\vec{f}_\text{ij}^{\textrm{DPD}} = \sum_{i \neq j} \left( \vec{f}^{\text{R}}_{\text{ij}} + \vec{f}^{\text{D}}_{\text{ij}}  + \vec{f}^{\text{C}}_{\text{ij}} \right) ,
\label{eqn:forces}
\end{equation}
where the random force $\vec{f}^{\text{R}}_{\text{ij}}$ is
\begin{equation}
\vec{f}^{\text{R}}_{\text{ij}} = \sigma\ w^{\text{R}}(r_{\text{ij}}) \ \theta_{\text{ij}} \ \hat{\vec{r}}_{\text{ij}} ~,
\end{equation}
the dissipative force $\vec{f}^{\text{D}}_{\text{ij}}$ is
\begin{equation}
\vec{f}^{\text{D}}_{\text{ij}} = - \gamma\ w^{\text{D}}(r_{ij})\ (\hat{\vec{r}}_{\text{ij}} \cdot \vec{v}_{\text{ij}} ) \ \hat{\vec{r}}_{\text{ij}} ~,
\end{equation}
and $\vec{f}^{\text{C}}_{\text{ij}}$ is a conservative force, which we define in
Eq.~\eqref{fc}. $\vec{r}_{\text{ij}}$ is the distance vector, $r_{\text{ij}}$ is the
distance, $\hat{\vec{r}}_{\text{ij}}$ is the unit vector along the direction of the distance
vector, and $\vec{v}_{\text{ij}}$ is the relative velocity.
$\theta_{ij}$ is a white noise with the following properties:
\begin{equation}
\left< \theta_{\text{ij}} \right> = 0
\end{equation}
\begin{equation}
\left< \theta_{\text{ij}}(t) \theta_{\text{kl}}(t') \right> = (\delta_{\text{ik}} \delta_{\text{jl}} +  \delta_{\text{il}} \delta_{\text{jk}}) \delta(t-t')
\end{equation}
where $\delta_{\text{ij}}$ is Kronecker's delta and $\langle \bullet \rangle$ denotes
averaging with respect to time.

The factors $\sigma$ and $\gamma$ characterize the strength of the random and dissipative
force, subject to the weight functions $w^{\text{R}}$ and $w^{\text{D}}$.
In the DPD simulation model, all the components of the force are short-ranged with a cutoff
distance $r_\textrm{cut}$. This is typical of particle-based simulation models and necessary
for the technique of domain decomposition (see Sec. \ref{sec:cell-system}).
Long-range forces, which must be taken into account for electrostatic or dipolar interaction
between colloids, are computed with dedicated routines in ESPResSo for the corresponding
simulation scenarios.

For DPD simulations---as it is the case for the Langevin thermostat---we fix the relation
between the intensity of the noise and the friction parameter by using the
fluctuation-dissipation theorem:\cite{Espanol1995statistical}
\begin{equation}
\label{eq:fdt}
\sigma^2 = 2 k_B T \gamma ~,
\end{equation}
where $k_B$ is Boltzmann's constant and $T$ is the temperature. In practice, we pick a value
for the temperature $T$ and for the friction $\gamma$, and the noise amplitude $\sigma$ is
set by ESPResSo's DPD thermostat according to Eq.~\eqref{eq:fdt}.
It is furthermore required to select the weight function $w^{\text{R}}$ and $w^{\text{D}}$
so that
\begin{equation}
\left[ w^{\text{R}}(r)\right] ^2 = w^{\text{D}} (r)
\end{equation} 
We chose $w^{\text{R}}$ as
\begin{equation}
w^{\text{R}}(r) = 1 - \frac{r}{r_{\text{cut}}} ~,
\end{equation}
which is a common choice.\cite{groot1997}

In DPD simulations, it is customary to use a soft repulsive interaction with amplitude $a_{ij}$ to account for conservative (C) forces $\vec{f}^{\text{C}}_{\text{ij}}$. We use
\begin{equation}\label{fc}
\vec{f}^{\text{C}}_{\text{ij}}= a_{\text{ij}} \left( 1 - \frac{r}{r_{\text{cut}}} \right) 
\end{equation}
for $r < r_{\text{cut}}$ and zero otherwise. This soft potential allows the use of larger time steps in the simulation.

\subsection{Green-Kubo techniques} \label{subsec:GK}
The simplest way to evaluate bulk properties in MD simulations is to use Green-Kubo relations. They allow bulk properties to be connected to macroscopic fluxes caused by thermal fluctuations. In the present study, we focus on the self-diffusion coefficient $D$ and the shear viscosity $\eta$. Thus we evaluate the fluxes of the particle velocities and the shear stress. The characteristic equation for the self-diffusion coefficient is
\begin{equation}
D = \frac{1}{3} \int_0^\infty \langle \vec{v}_i(0) \vec{v}_i(t+\tau) \rangle_{|t} d\tau
\label{eqn:GK_diff}
\end{equation}
where $\vec{v}$ represents the velocity of an individual particle $i$ in three dimensions. The
angular brackets $\langle \ \rangle_{|t}$ represent the ensemble average over all lag times present in the
simulation. We use $\tau$ as the lag time.
For the shear viscosity in the unsheared case, we use
\begin{equation}
\label{eq:gk}
\eta = \frac{V}{k_BT} \int_0^\infty \langle \sigma_{xy}(0) \sigma_{xy}(t+\tau) \rangle_{|t} d\tau
\end{equation}
where $k_B T$ is the thermal energy, $V$ is the box volume, and $\sigma_{xy}$ is the off diagonal element of the
instantaneous virial stress tensor, also known as Irving-Kirkwood stress tensor,
\begin{equation}
\label{eq:stress}
\sigma_{k,l} = \frac{\sum_i m_i v_i^{(k)} v_i^{(l)}}{V} + \frac{\sum_{j>i} F_{ij}^{(k)} r_{ij}^{(l)} }{V}
\end{equation}
where $k$ and $l \in [x, y, z]$ indicate the dimension of the coordinate.

\subsection{Brownian motion} \label{subsec:MSD}

The migration of the fluid particles with time is evaluated using the mean-square
displacement (MSD)
\begin{equation}
\text{MSD}(\tau) = \langle \left( \vec{x}(t+\tau) - \vec{x}(t) \right)^2 \rangle
\label{eq:pure_msd}
\end{equation}
where $\vec{x}$ describes the position of a particle in three dimensions. In the absence of the shear
flow, the MSD allows the diffusion coefficient to be calculated using the relation
$6Dt = \text{MSD}(t)$.
The computation of the mean square displacement is also relevant in shearing and non-steady state regimes.
Furthermore, it can be evaluated for several directions independently allowing more detailed
insight. This is of particular interest for the simulations with shear flow.\cite{Foister1980}
Several predictions can be made for the diffusion of Brownian particles under shear flow. The MSD of our DPD fluid particles should follow the prediction for Brownian particles. In steady shear flow, the displacement of particles along the shear direction $x$ is given by \cite{Orihara2011}
\begin{equation}
\langle \left( x(t) - x(0) - \dot \gamma z(0) t \right)^2 \rangle = 2 Dt  \left[1 + \frac{1}{3}(\dot\gamma t)^2  \right]
\label{eq:msd}
\end{equation}
with a cubic dependence of time, indicating an enhanced diffusion due to the particles
migrating through regions with different shear velocities \cite{Foister1980}.
The term $-\dot\gamma z(0) t$ on the left-hand side of Eq.~\eqref{eq:msd} removes the mean
horizontal drift that stems from the initial velocity of the particle. By measuring the MSD
in this manner, it is possible to observe the cubic dependency also observed by Orihara and
Takikawa~\cite{Orihara2011}.

For oscillatory shear flow, the MSD in the shearing direction $x$ follows the relation
$\langle \Delta x(t)^2 \rangle = 2 D_{\text{eff}}t$ with
\begin{equation}
D_{\text{eff}} = D \left[1 + \frac{\gamma_0^2}{2} \left(2 \sin^2\Phi +1 \right) \right]
\label{eq:msd_osc}
\end{equation}
where $\Phi$ represents the phase and $\gamma_0$ the deformation amplitude\cite{Takikawa2012}.
The motion of the Brownian particles in an oscillatory shear flow combines a periodic and a
diffusive component. As Takikawa and Orihara\cite{Takikawa2012}, we evaluate the position of
the particles in a stroboscopic manner---that is with a time interval that is a multiple of
the forced oscillation period---so that the resulting motion appears as purely diffusive
with the modified diffusion coefficient $D_{\text{eff}}$ that depends on the amplitude of
the shear flow and on the phase of the oscillatory movement.
\label{eq:osc}

\subsection{Calculation of correlations}
\label{sec:correl}

We rely on two different procedures to compute formulas of the form
\begin{equation}
\label{eq:corr}
\langle X(t_i) X(t_j) \rangle
\end{equation}
found in Eqs.~\eqref{eqn:GK_diff} and \eqref{eq:pure_msd}.
A logarithmic correlator is available in ESPResSo for the set of built-in observables. Such
a correlator samples the term $X(t_i) X(t_j)$ for fixed time differences
$[0, M^m\Delta t, 2M^m\Delta t, \dots, N\cdot M^m\Delta t]$, for consecutive values of the
exponent $m$, taking the form of blocks having time intervals that increase by a factor $M$
between successive blocks.
Storing lag times that are $M$ times larger implies the addition of $N$ samples instead of
$M$ times more samples: Storing samples up to a time lag
$\tau_{max} = N\cdot M^{m_{max}}\Delta t$ requires $m_{max} N$, which is
$\mathcal{O}(\log \tau_{max})$ hence the name logarithmic correlator. This technique, which
is useful when the number of samples would otherwise exceed the available memory, is
presented in the book by Frenkel and Smit~\cite{Frenkel2002}.

Another technique is the Fast Correlation Algorithm (FCA) that relies on Fourier
transforms~\cite{nmoldyn_1995} to speed up the computation. We use the implementation
provided by the Python package tidynamics~\cite{tidynamics_2018} for autocorrelation and
mean-square displacements. We refer to this method as a linear correlator as it requires
data samples linearly spaced in time.
To use the FCA method, we store the variables of interest to disk (the position for the mean
squared displacement or selected components of the stress tensor for the viscosity).

The logarithmic correlator and the FCA only differ in their statistical sampling. The FCA
method is equivalent to the computation of the pairwise correlation for all time intervals available
and provides the same results, up to rounding errors, as computing the correlations with a
naive $\mathcal{O}(N_{samples}^2)$ loop, where $N_{samples}$ is the total number of sample
times.

For the computation of Eq.~\eqref{eq:msd_osc}, we do not perform an average over time. The
correlation is not of the form \eqref{eq:corr} and the two techniques presented above do not
apply.

\section{Results and discussion}
\label{sec:results}

We carried out all of our simulations with 10,000 particles and densities of $\rho = [3, 4, 5, 6, 7]$. We used a strength of the repulsive parameter from $a_{ij} = 0$, i.e. no repulsive force, up to $a_{ij}=175$. These parameters are similar to the ones used by Zohravi {\em et al.}~\cite{Zohravi2018} as this was the most complete study concerning the influence of the density $\rho$ and the strength of the conservative interaction parameter $a_{ij}$ on the shear viscosity, thus providing a good
reference point to benchmark our method. The temperature is set to $k_BT=1$ and the friction constant $\gamma$ to 4.5. All simulations use a time step of $\Delta t = 0.005$.
The results are available in full in the analysis notebooks, see appendix \ref{sec:reproc} for details.

\subsection{Self-diffusion coefficient and viscosity of DPD fluids}

In this section, we use the mean square displacement, Green-Kubo techniques, and Lees Edwards boundary conditions to evaluate the equilibrium and non-equilibrium properties of the DPD fluid. First, we start by measuring the self-diffusion coefficient $D$ via the mean square displacement and then we investigate the shear-viscosity $\eta$ and its various contributions using the two other mentioned methods. Simulations using Green-Kubo were conducted at quiescent conditions whereas samples under shear used the LEbc method.

\subsubsection{Self-diffusion coefficient}
\begin{figure}[htbp]
\begin{center}
\includegraphics[width=0.5\textwidth]{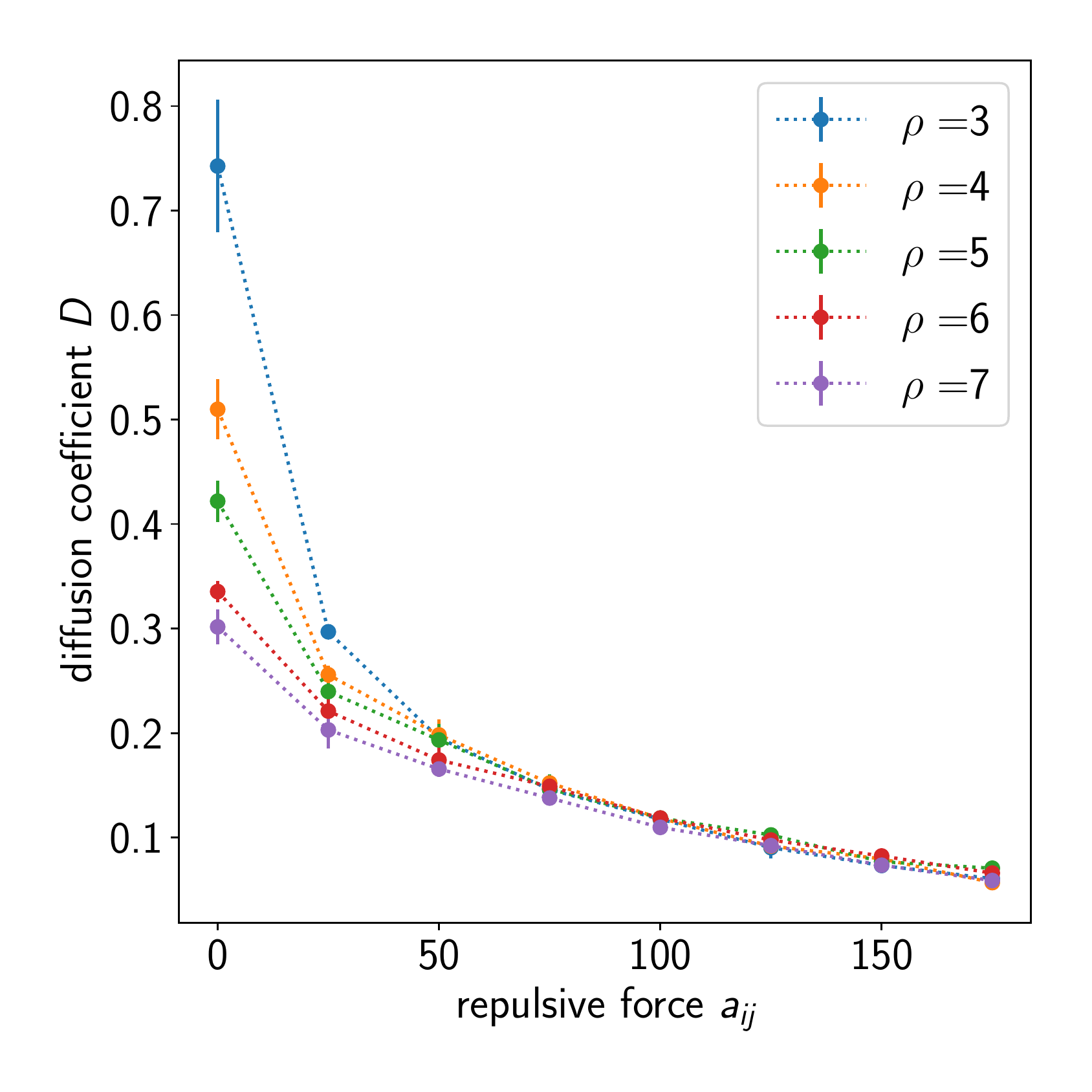}
\caption{\bf{The self-diffusion coefficient of a quiescent DPD fluid in dependence of the repulsive force parameter $a_{ij} $} for the densities $\rho = [3, 4, 5, 6, 7]$ and $k_B T =1.0$. The lines connecting the symbols are to guide the eye.}
\label{fig:diff_coeff_quiescent}
\end{center}
\end{figure}

We show the diffusion coefficient that was obtained from the mean square displacement, equation \eqref{eq:pure_msd}, of 10,000 particles at $k_BT = 1.0$ in Figure~\ref{fig:diff_coeff_quiescent}. Each individual trajectory of the particle was correlated using the logarithmic correlator from ESPResSo and subsequent fitting of $\mathrm{MSD} = 6 D t$. This results in 10,000 individual diffusion coefficients per simulation run. The results, shown in Figure \ref{fig:diff_coeff_quiescent}, are obtained from an average of three individual quiescent runs.
Our results show that the diffusion coefficient decreases with an increase in repulsive strength as is expected since particles are hindered in their movement by the other surrounding particles and their repulsive interaction.
By increasing either the number density of the particles or their repulsive strength, the overlap between the repulsive spheres increases, hence, the movement is hindered through ``caging" particles with other DPD particles due to the higher number of possible interaction partners or the higher interaction strength. This leads to the lower effective diffusion coefficients with increasing density and increasing repulsive strength shown in Figure \ref{fig:diff_coeff_quiescent}. The diffusion coefficient reduces monotonically with increasing repulsive strength for all number densities. Furthermore, the figure shows that diffusion coefficients decrease monotonically as a function of particle density. This effect is present over the entire range of repulsive forces.

\subsubsection{Viscosity}
\begin{figure*}[htbp]
\begin{center}
\includegraphics[width=1.0\textwidth]{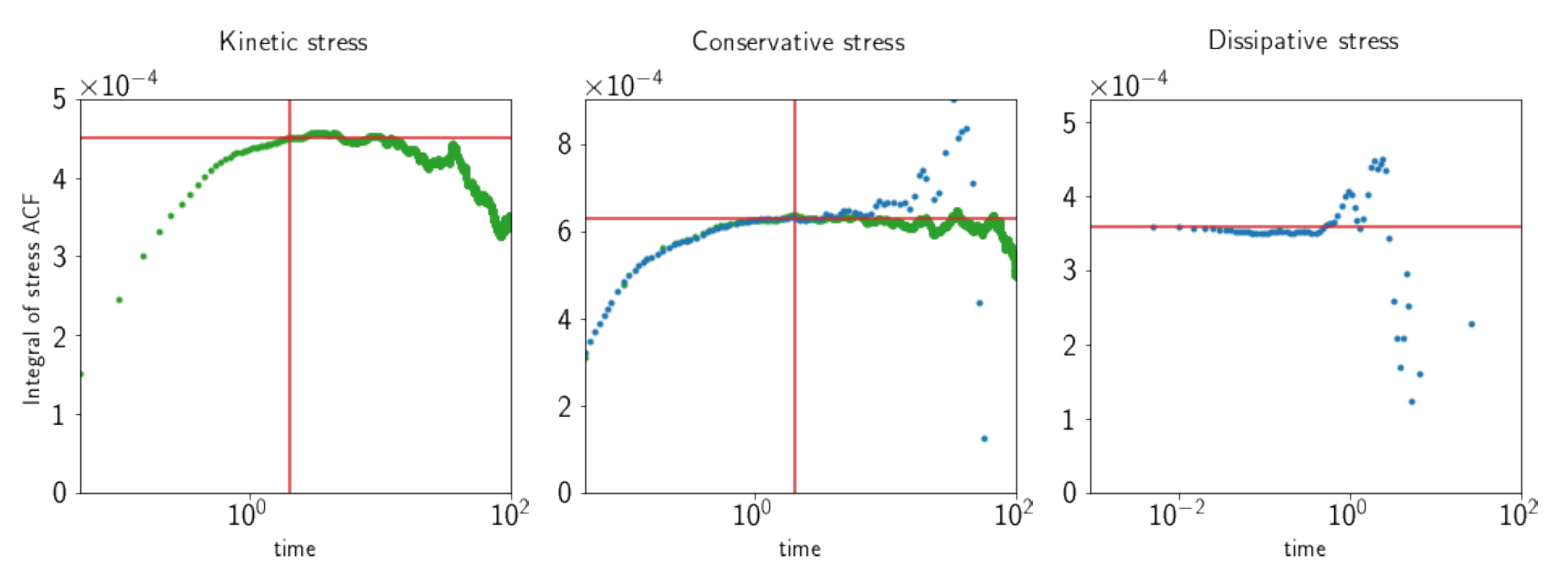}
\caption{\bf{Integral of the autocorrelation (ACF) of the DPD fluid ($\rho = 6.0$ and $a_{ij} = 25$) as measured by the Green Kubo method. Data obtained using a linear correlator are shown in green and a logarithmic correlator in blue. Red lines indicate the time cutoff and the corresponding plateau value.}}
\label{fig:acf_convergence}
\end{center}
\end{figure*}

The viscosity of the DPD fluid, on the one hand, is connected to the inter-particle forces that are described in equation~\eqref{eqn:forces} and on the other hand to the overall momentum of the particles that results in the kinetic contribution. The contributions can be further divided up into the viscosity based on the random force $\vec{f}^{\text{R}}$, the dissipative force $\vec{f}^{\text{D}}$, the conservative force $\vec{f}^{\text{C}}$. The sum of the kinetic and the conservative viscosity is referred to as the total viscosity $\vec{f}^{\text{T}}$. We measure the viscosity using the Green-Kubo method as well as the Lees Edwards method, which measures the instantaneous stress at the ``wall". That means that we include the noise in these measurements as it generates a non-negligible contribution to the integral of the autocorrelation.
We measured the viscosity by two methods: First, by the Green-Kubo formula~\eqref{eq:gk} and, hence, by measuring the stress fluctuations in quiescent simulations and second, by directly measuring the stress in a fluid sheared with the Lees-Edwards method which will be explained in the next part of the paper.
Here, we start by discussing the results of the quiescent simulations. In order to fully describe our methods, we first show how we obtained these results.
Green Kubo results are calculated with the logarithmic correlator of Espresso. We also sampled the data in a trajectory file at linear time intervals and used the \texttt{acf} method of tidynamics. The online correlator collects data up to $\tau = 100,000$. Following the initial warm up of 1,000,000 integration steps, we run the simulation for a total 500,000 time steps of $\Delta t = 0.005$. We then plot the integral of the autocorrelation function and choose a uniform cut-off for all the simulation in order to avoid any bias between the iterations.
The convergence of the autocorrelation function as needed by the Green Kubo method is illustrated in Figure~\ref{fig:acf_convergence}, where vertical red lines show the time cutoff and horizontal red lines show the corresponding plateau value.

Depending on the contribution, we show the two possible correlation methods. For the kinetic component, a linear correlator was used (shown in green in the left plot of Figure~\ref{fig:acf_convergence}). This linear correlator was necessary because the kinetic component cannot be extracted from the simulation package directly. A distinct plateau beginning at $t=2.0$ can be found in this data. For the conservative stress, Figure 4 center, we used a linear (green) as well a logarithmic (blue) correlator, to illustrate that the differences between them are only due to the sampling. Once more, a distinct plateau beginning at the cut-off is visible. The dissipative part of the viscosity as shown in the right plot of Figure~\ref{fig:acf_convergence} could only be evaluated using the logarithmic correlator. The dominant part of this viscosity contribution originates in the delta peak around $\tau = 0$ of the autocorrelation due to the sampling of the random noise. The fluctuating contribution to the autocorrelation function are very small. We decided to cut off the integral at this very initial point where the noise starts to dominate.

\begin{figure}[htbp]
\begin{center}
\includegraphics[width=0.5\textwidth]{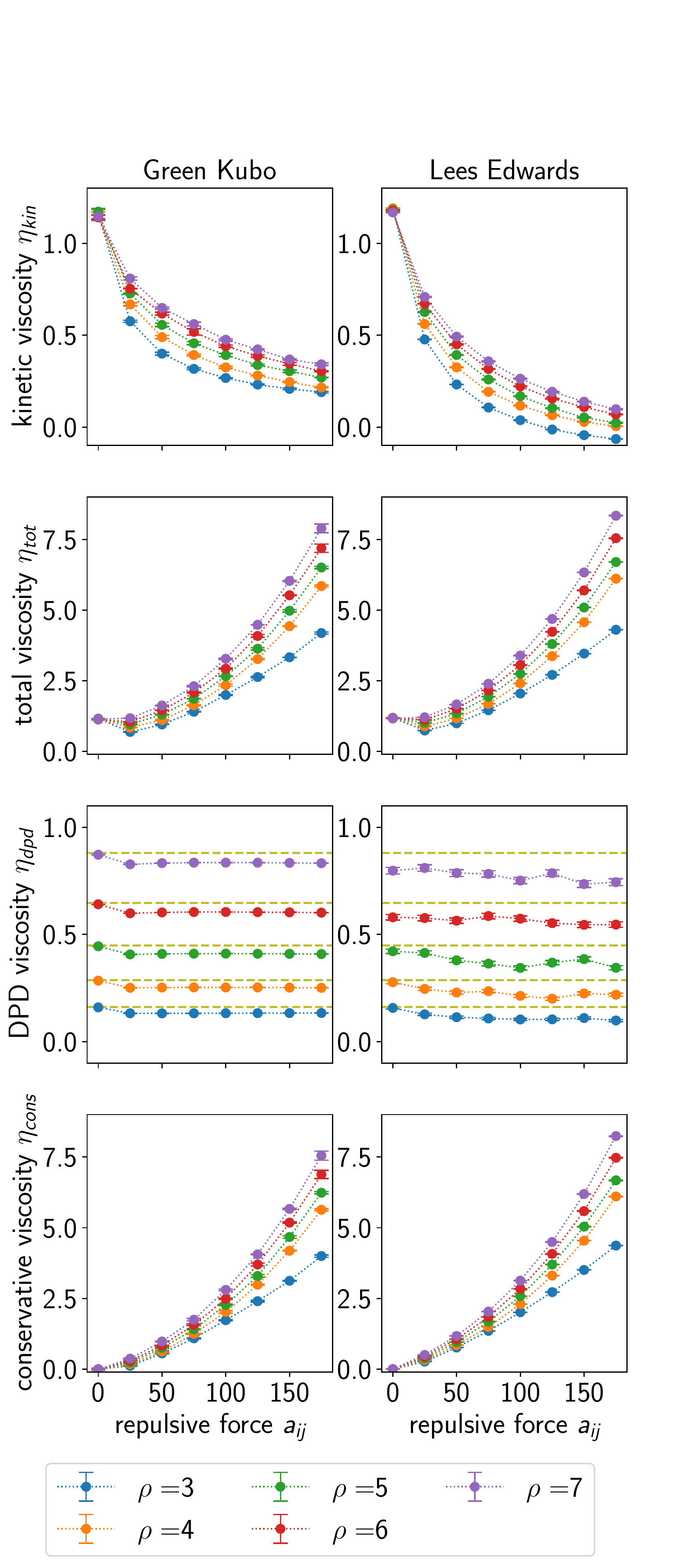}
\caption{\bf{The viscosity of a DPD fluid for various repulsive force parameters $a_{ij} $} for the densities $\rho = [3, 4, 5, 6, 7]$, at $k_b T = 1.0$. The left column shows the results of the quiescent Green Kubo analysis and the right column shows the same values as obtained with shearing Lees Edwards boundary conditions. The dashed line shown in the DPD viscosity represent the approximation found by Groot and Warren \cite{groot1997}. The dotted lines are shown to guide the eye.}
\label{fig:viscosity_comparisson}
\end{center}
\end{figure}

The results for the Lees Edwards experiments were obtained from three simulation runs per
data point. The warm up of our fluid consists of 100,000 integration steps. After this warm
up, we turn on the shear flow with $\dot \gamma = 1.0$ and re-equilibrate 2 runs for another
100,000 integration steps. As a test, one run was re-equilibrated for 500.000 integration
steps, but showed the same results. The resulting equilibration time is larger than what is
necessary to reach the linear regime in the Navier-Stokes equation, using there an
approximation for the fluid viscosity.
Then we start the data recording and obtain 200,000 stress values, 100 integration steps apart from each other (20,000,000 integration steps in total). We applied the blocking method \cite{flyvbjerg1989} to obtain mean and standard deviation of this data and present the results in the right column of Figure \ref{fig:viscosity_comparisson} using the pyblock Python module \cite{pyblock}. For this analysis, we only used the last $2^{17} = 131,072$ measurement values in order to ensure a steady shear profile is obtained. The usage of more or fewer data points, e.g. the last $2^{16}$ or 170,000, did not change the results, Therefore, we are confident with the assumption that a stationary regime was obtained. We also chose the number of sampling points as a power of 2 because it is most efficient to apply the blocking method on such a data set. 

In Figure~\ref{fig:viscosity_comparisson}, we show the collected results from all simulations. The left column shows the quiescent results from the Green Kubo method and the right column shows the shear results from the experiments using the Lees Edwards method. We also show a superimposed view of this data in the analysis notebook, part of the supplementary information to this paper. 
The kinetic viscosity decreases with increasing repulsive strength for both methods. This is in agreement with the reduced diffusion constant of the particles (Figure~\ref{fig:diff_coeff_quiescent}). Samples with a lower density $\rho$ have a lower kinetic viscosity, except for the case of $a_{ij} = 0.0$ (no repulsion). Here, the kinetic viscosity is the highest for all samples with a value of around $\eta_{kin} = 1.2$. The same trend can be observed for viscosity measured via the Lees Edwards technique. However, the reduction in the kinetic viscosity is more pronounced; the quiescent method always showing higher viscosities than for Lees Edwards at high repulsive forces.
Since the total viscosity consists of the contribution from the kinetic part and from the conservative part that means it takes the contributions of the increasing repulsive strength into account. The results are consistent with the case of no repulsive force present where all samples show the same kinetic and total viscosity.
We have noticed small negative values for the Lees-Edwards result, for $\rho=3$ and the
higher values of $a_{ij}$, but the fluctuations of the kinetic stress are much larger than
the actual value which makes it a difficult measurement.
While there is a difference between the kinetic component of the viscosity between the
quiescent and shearing samples for some values of the parameters, the similarity between the
Green Kubo and Lees Edwards results is remarkable overall. There is only a maximal
difference of $\frac{\Delta \eta}{\eta} = \frac{\eta_{le} - \eta_{gk}}{\eta_{gk}} = 0.07$
between the two methods.

In our experiments, we also directly measure the dissipative part of the viscosity and compare these values to the prediction by Groot and Warren \cite{groot1997}.
Figure \ref{fig:viscosity_comparisson} shows that, for no repulsive force present, the data points from Green Kubo lie exactly on the predicted line whereas all points seem to be slightly shifted downwards but remain on a constant value for higher repulsive forces. For the measurements obtained by the Lees Edwards technique, a bigger separation between the theoretical prediction and the obtained values can be observed. The values for an absent repulsive force also already deviate from the predicted values. A possible reason for this discrepancy could be caused by the linear interpolation of the velocity differences between the particles at the boundary. This might underestimate the real velocity difference and, hence, also the real stress caused by the relative movement. Furthermore, the introduction of additional energy via the shear could change the system in a way that makes is effectively different from the quiescent one. 

The conservative viscosity shows a similar trend as the total viscosity. It starts from $\eta_{\textit{cons}} = 0.0$ in the cases without conservative force and increases monotonically thereafter. The higher the density of a sample, the steeper is this increase. Error bars in this plot are based on the sum of errors from the kinetic viscosity and the total one. 
Overall, our experiments clearly show an agreement between both the static and the dynamic
measuring technique, even though the trends and numerical values for the Lees Edwards
measurements are less obvious and show a larger error.
It would be useful in later work to study in more detail the dynamics of the sheared fluid
to provide a better assessment of the difference between the quiescent fluid and the sheared
fluid.

\subsection{Flow profile}
\label{subsec:flowprofile}

\begin{figure}[htbp]
\begin{center}
\includegraphics[width=0.5\textwidth]{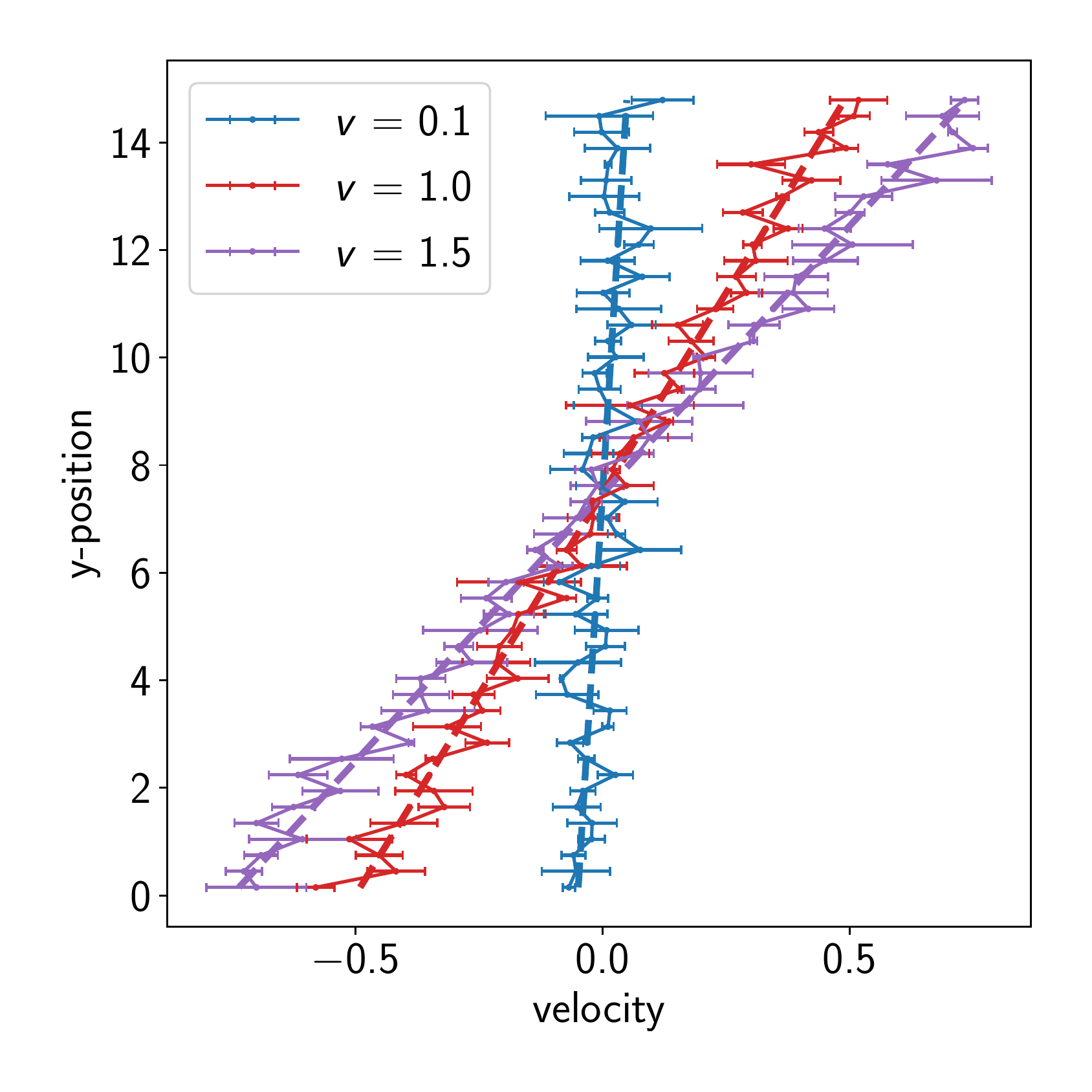}
\caption{\bf{Height dependent velocity profile of the DPD fluid for 3 different shear velocities $v = [0.1, 1.0, 1.5]$. The expected linear shear profile is shown as a dashed line.}}
\label{fig:flowprofile}
\end{center}
\end{figure}

We perform simulations for a DPD fluid with $N=10000$ particles, a number density
$\rho = 3$, a friction coefficient $\gamma = 4.5$, a cut-off radius $r_{\text{cut}} = 1.0$ and $F_{\text{max}} = 25.0$. The shear velocities in our simulations were $v = 0.1$, $1.0$, and $1.5$ and represent the velocity added to the particle when it crosses the lower boundary or, respectively, subtracted from the particle when it crosses the upper boundary. The established shear gradient leads to flow velocities of $- v/2$ at the bottom of the simulation box and of $+v/2$ at the top of the simulation box.
We investigated the height dependence of the flow velocity in the gradient direction to check if the flow profile was properly equilibrated and uniform across the box. For this purpose we divided the box in 50 horizontal slabs, oriented along the gradient direction and determined the average velocity of the DPD fluid particles in each slab after a start-up time of 1000 $t=50,000$. We show the average and standard deviation based on three different, independent snapshots for three different shear velocities in Figure \ref{fig:flowprofile}. The expected flow linear flow profile is also included as dashed lines for comparison.

The resulting shear gradient is linear and in good agreement with the expected shape. Furthermore, an increase in the number of bins to improve the sampling resolution along the box height did not have an impact on the linearity or slope. We find that our implementation does not show any discontinuity or spurious flows in the shear profile near the simulation boundary. Therefore, the correction proposed by Chatterjee of omitting dissipative forces at the boundary is not required here and such discontinuities are not inherent to Lees-Edwards boundary conditions. We conclude that the corrections suggested by Chatterjee~\cite{Chatterjee2007} are not necessary and confirm the previous research by Leimkuhler and Shang~\cite{Leimkuhler2016}.

\subsection{Brownian motion with shear flow}

The analysis of the mean-square displacement (MSD) enables the identification of Brownian
motion by the linear dependence of the MSD on time. Whereas the viscosity of the DPD fluid is of Newtonian character, the MSD is influenced by the shear.
In sheared systems, there exists a cubic-in-time contribution (Eq.~\eqref{eq:msd}) to the MSD that was observed
experimentally for polystyrene spheres by Orihara and Takikawa~\cite{Orihara2011}.

In computer simulations using the Lees-Edwards method, the study of diffusion depends on the
ability to reconstruct the physical trajectories of the particles even though they experience ``jumps'' when crossing the boundaries.
As in the case of periodic boundary conditions, the coordinates are wrapped in the primary
simulation box. Instead of using the plain unwrapped coordinates, based on the number of
jumps in each direction, we use the accumulated offset defined in
Eq.~\eqref{eq:accumulated-offset} to obtain physically consistent trajectories.
The study of Brownian motion thus serves as an extra verification of the correctness of our
implementation.
Once more we perform simulations for a DPD fluid with with the parameters mentioned in subsection \ref{subsec:flowprofile}. A repulsive force of $F_{\text{max}} = 25.0$ for the continuous shear simulations and $F_{\text{max}} = 5.0$ for the oscillatory shear simulations was used. We used a lower value for $F_\text{max}$ in the oscillatory case to obtain a higher diffusivity for the DPD particles and hence a better signal to noise ratio. The effective diffusion coefficient $D_{\textit{eff}}$ for oscillatory shear depends on the strain and phase of the movement. As these are both values restricted by the simulation (e.g. the time of the shear wave to travel through the box) we had to enhance the diffusion of the particles to show the effect in an illustrative way.

\subsubsection{Continuous shear}

We chose the same simulation conditions as reported in subsection \ref{subsec:flowprofile}
and five different shear velocities between $v = 0.1$ and $v = 1.5$ resulting in shear rates
ranging from $\dot \gamma \approx 0.003$ and $\dot \gamma \approx 0.05$. The mean-squared
displacement (MSD) of the particles was measured after equilibration of the shear flow.

\begin{figure}[htbp]
\begin{center}
\includegraphics[width=0.5\textwidth]{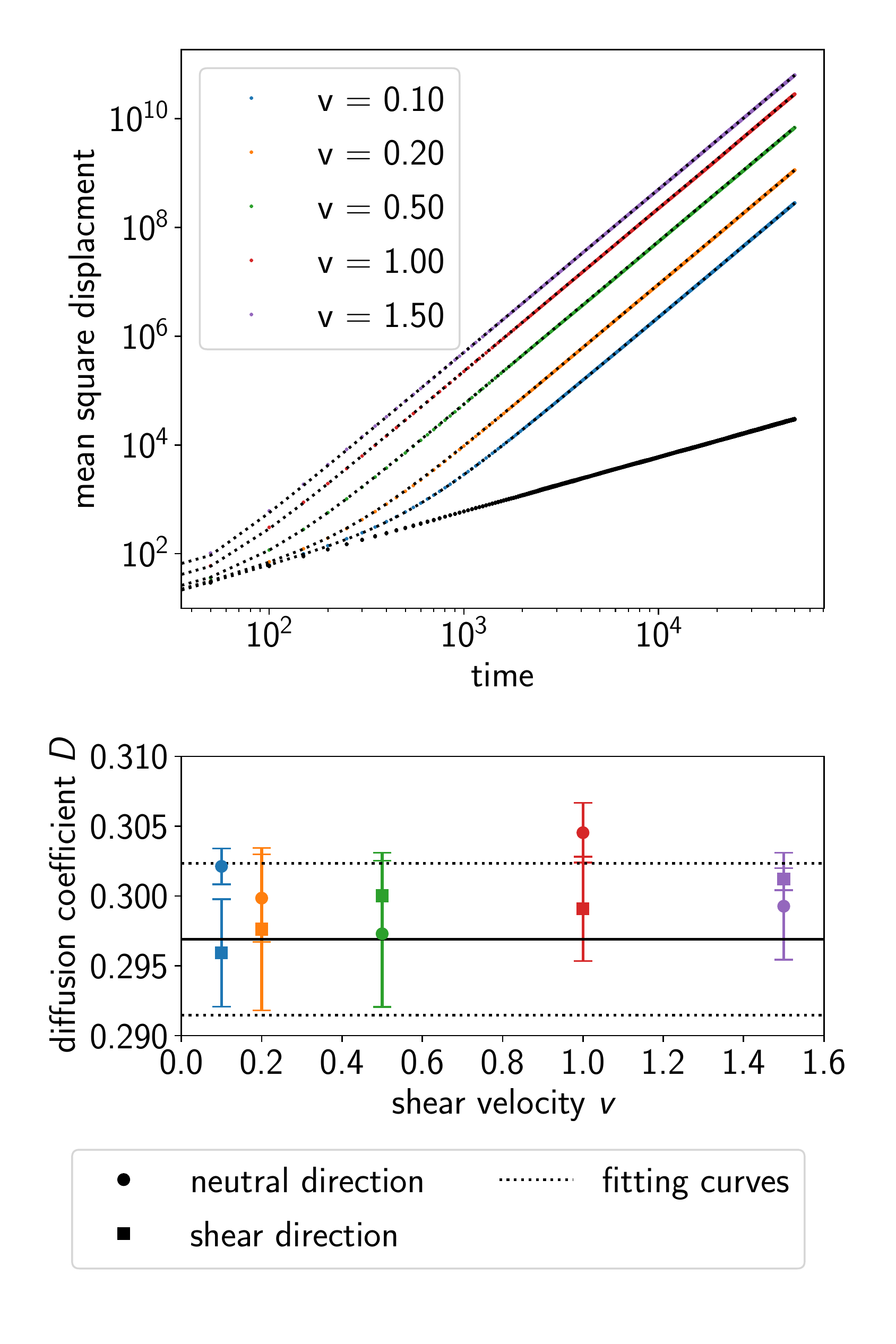}
\caption{\bf{Results for the MSD and diffusion coefficient $D$ in continuous shear. The upper panel shows the development of the MSD over time for the neutral direction (black), which is linear, and for the shearing direction, where the leading term is cubic at large times, for 5 different shear rates $\dot \gamma$. To obtain the diffusion coefficients in the lower panel, we fitted the MSD in the neutral direction (with a linear function) and in the shear direction with Eq.~\eqref{eq:msd} with a set value for $\dot\gamma$. The fitted curves, in black, match the simulation data. In the lower panel, the black solid line indicates the value of $D$ obtain from the quiescent simulation (see Fig.~\ref{fig:diff_coeff_quiescent}), with the dotted lines at $\pm$ one standard deviation. The round (square) symbols show the value of $D$ in the neutral (shear) direction.}}
\label{fig:msd_conti}
\end{center}
\end{figure}

Figure \ref{fig:msd_conti} shows the MSD for the neutral and the shearing direction for one
example per shear rate. The colored curves show the actual measurement data while the black
dotted lines show fitted curves to this data. We comment on the fitting procedure and the
relation between the curves in the caption.
Following the relations presented in subsection \ref{subsec:MSD} the linear in time
character of the MSD in the neutral direction is unaffected by the shearing. The MSD in the
shearing direction shows a gradual transition to a cubic dependency as a function of time,
as we expect from equation~\eqref{eq:msd}. The data and the superimpose and are
undistinguishable in the figure.
The lower part of Figure~\ref{fig:msd_conti} shows the measured diffusion coefficient $D$
and standard deviation for the neutral and shearing direction as determined by three
independent runs. We obtain these values by fitting the theoretical expressions of
subsection \ref{subsec:MSD} to the measured values. The ratio of the diffusion coefficient
in the neutral and vorticity direction has a maximum value of around 2.5\%.
We thus confirm numerically the validity of Eq.~\eqref{eq:msd}. This measurement, in a
particle-based simulation using the Lees-Edwards method, is only possible thanks to the
observation of the reconstructed trajectories based on Eq.~\ref{eq:accumulated-offset}.
The observation of the diffusion of simple particles in shear flow only depends on the
diffusion coefficient and on the shear rate, so that the same analysis holds for the
experiments of Orihara and Takikawa~\cite{Orihara2011} and our simulations.

\subsubsection{Oscillatory shear}

We use the same settings as for the continuous shear flow experiments but a reduced conservative force of $F_{\text{max}} = 5.0$, in order to enhance the diffusion, and an oscillation period of 500. The diffusion coefficient in the neutral direction during the oscillatory flow is $D = 0.61 \pm 0.01$. We then plot the expected effective diffusion coefficient $D_{\text{eff}}$ following equation~\eqref{eq:msd_osc}. The results shown were obtained from fits to 299 periods of oscillatory shear. This way, we can slide a window over the trajectories to obtain results for different phases. Fittings of the MSDs were cut off at $\tau = 10^4$ as the MSD at larger times is the result of too few averaging points.

\begin{figure}[h!]
\begin{center}
\includegraphics[width=0.5\textwidth]{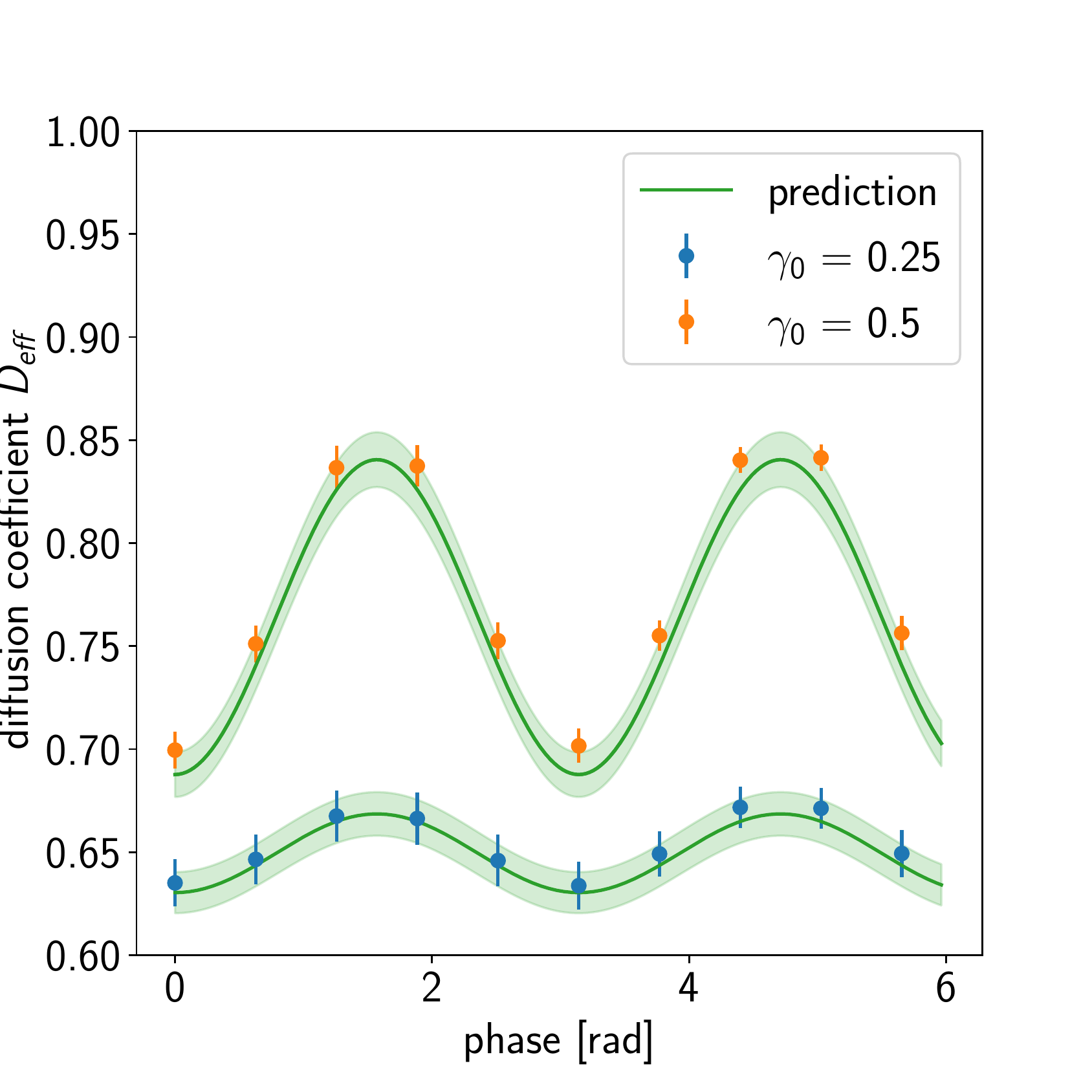}
\includegraphics[width=0.5\textwidth]{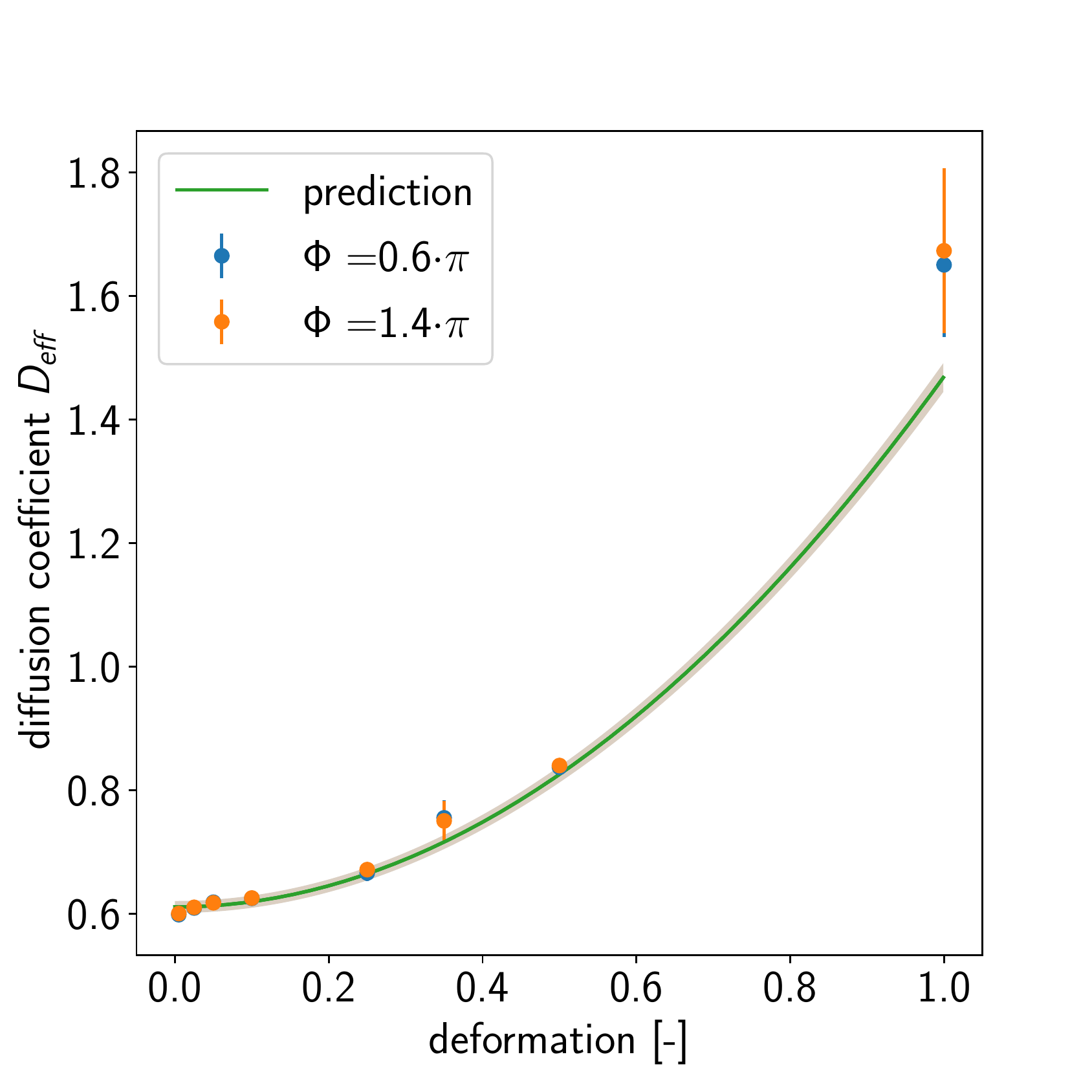}
\caption{\bf{Results for the MSD and diffusion coefficient $D_{\text{eff}}$ in oscillatory shear. The upper part illustrates the phase dependent diffusion coefficient for two different strain amplitudes $\gamma_0 = 0.25$ and $\gamma_0 = 0.5$. The lower part shows the strain dependence of the diffusion coefficient $D_{\text{eff}}$.} for two phases $\Phi$. All predictions are based on Equation \ref{eq:msd_osc} and the diffusion coefficient measured in the neutral direction.}
\label{fig:msd_osc}
\end{center}
\end{figure}

For oscillatory shear, we can obtain a phase dependent diffusion coefficient, as shown in Figure \ref{fig:msd_osc} for two different strain amplitudes $\gamma_0$.  Furthermore, we show the strain dependence of $D_{\text{eff}}$ at two fixed phases $\phi$ which follows a squared relationship with the strain. Both results indicate the correct handling of jumps across a boundary and the correct handling of interactions. For large strains, we can observe a deviation of the measured $D_{\text{eff}}$ that is significantly higher.  This deviation can be explained by the high velocity at the boundaries. Since the shear velocity is higher in this case.

Hence, Lees Edwards boundary conditions are indeed translationally invariant and do not require any special modification at the boundary in order to avoid a spurious discontinuities in the flow. As long as the thermalization and velocity difference is calculated correctly by taking the shear velocity into account one can even model shear flow phenomena with a shear velocity that is changing over time.

\section{Conclusions}
\label{sec:conclusions}

We have designed and implemented the method proposed by Lees and Edwards in 1972 for the
simulation of linear shear flows in Molecular Dynamics.
We provide in section \ref{sec:implementation} the information for the practical
implementation in the simulation software, specifically on the distance function, the
velocity difference function, the cell system, and the storage of the trajectory offset,
that will be useful as a starting point for other scientists. In addition, our code is
available publicly under an open-source license.

We demonstrated the Lees-Edwards method with a dissipative particle dynamics (DPD) fluid, a
common choice in mesoscopic fluid simulations, to obtain a linear velocity profile. We find
a good agreement between the equilibrium and non-equilibrium properties of DPD fluids as
evaluated by Green-Kubo, for quiescent experiments, and by Lees-Edwards boundary conditions
experiments under shear flow.
Here, further work could be interesting to study the low shear-rate limit with the
Lees-Edwards method. While requiring longer simulation runs, this should shed light on the
remaining difference in numerical value seen in the comparison of
Fig.~\ref{fig:viscosity_comparisson}.
Next, we were able to reconstruct continuous trajectories from the shear simulations, as if
the system was infinitely extended, as is typically done for periodic simulation boxes.
We observe the diffusion of particles with the mean square displacement and diffusion
coefficient in equilibrium as well as in non-equilibrium situations, using then the
reconstructed trajectories. We recover the predicted enhanced diffusion of Brownian
particles in shear flow, which would be impossible to do without the quantity
$x_{\text{part, LE}}$ defined in Eq.~\eqref{eq:accumulated-offset}. These results are of
special interest as they allow for a direct comparison to the experiment of Orihara and
Takikawa under steady shear~\cite{Orihara2011} and to the one of Takikawa and Orihara under
oscillatory shear~\cite{Takikawa2012}.
As the results depend only on the diffusion coefficient and on the shear rate, they are
promising for in-silico preparatory work for other types of colloids such as non-spherical
colloids or polymers.

Our work opens up new possibilities to conduct numerical experiments involving simulations
that require an explicit solvent undergoing shear flow within the convenient simulation
package ESPResSo.
We confirm the results of Leimkuhler and Shang~\cite{Leimkuhler2016} that the combination of
DPD and Lees-Edwards yields a translationally invariant system, which ensures a sound basis
for further research with this simulation setup.

Other methods have been devised to simulate the motion of particles in shear flow, such as
the combination of the Lees-Edwards method with the ``Smoothed Profile Method''
(SPM)~\cite{kobayashi2011}.
It is possible, instead, to use DPD for representing the fluid in such applications. This
can for instance be useful in order to control the solvent quality of sheared polymer
solutions.  Another prospective use case is the yielding of gels where periodic boundaries
are necessary to capture the macroscopic behavior of extended gel systems. Highly localized
restructuring can lead to feedback between the applied shear deformation and network
structure that would not be captured by other methods. This work also allows us to capture
the shear-induced orientation of, e.g., soft particles or liquid crystals where many
neighboring interactions must be considered.

\begin{acknowledgments}
We acknowledge funding of the Research Foundation - Flanders (FWO) Odysseus Program (grant
agreement number G0H9518N) and from the International Fine Particle Research Institute
(IFPRI).
Pierre de Buyl was a postdoctoral fellow of the Research Foundation-Flanders (FWO) while
preparing most of this work.
The resources and services used in this work were provided by the VSC (Flemish
Supercomputer Center), funded by the Research Foundation - Flanders (FWO) and the Flemish
Government.
\end{acknowledgments}

\appendix

\section{Computational reproducibility}
\label{sec:reproc}

We perform the simulation with the package ESPResSo~\cite{Arnold2013a, Arnold2013b,
  Limbach2006, Weik2019}.
We based our work on version 4.0 of ESPResSo and took care to minimize the number of
locations modified.
Our modifications to ESPResSo are available on Zenodo~\cite{bindgen_sebastian_2021_4627017}.
ESPResSo is a C++ package with Python bindings, so that a simulation consists in a Python
program that configures and executes the algorithm from the C++ ``core''.
For dumping trajectories, we used the H5MD~\cite{h5md_cpc_2014} writer of ESPResSo. H5MD is
a HDF5-based specification for molecular simulation data.

We collected all the parameters, simulation programs, and analysis notebooks (jupyter
notebooks \url{http://jupyter.org/}) in a dedicated repository, also archived on Zenodo, for
reproducibility purposes~\cite{sebastian_bindgen_2021_4719091}.
We use NumPy~\cite{numpy_2020} for basic numerical operations, SciPy~\cite{scipy_2020}
for numerical integration and curve fitting, matplotlib~\cite{matplotlib_2007} for the
figures, h5py~\cite{collette_python_hdf5_2013} to read HDF5 files,
tidynamics~\cite{tidynamics_2018} to compute correlations, and pyblock~\cite{pyblock} for
the block analysis.

\section*{Data Availability}

The data that support the findings of this study are openly available in Zenodo,
Ref.~\onlinecite{bindgen_sebastian_2021_4627017} for the ESPResSo package including our
Lees-Edwards implementation and Ref.~\onlinecite{sebastian_bindgen_2021_4719091} for the parameter
and analysis files.

\bibliography{bibliography}

\end{document}